\documentclass[prd,preprint,superscriptaddress,nofootinbib,longbibliography,aps]{revtex4-1}
\usepackage[utf8]{inputenc}
\usepackage{comment}
\usepackage{soul}
\usepackage{graphicx}
\usepackage{longtable}
\usepackage{amsmath}
\usepackage{color} 
\usepackage{amssymb}
\usepackage{subfigure}
\usepackage{tabularx}
\usepackage{microtype}
 \usepackage[linktoc=all]{hyperref}
 \hypersetup{
    colorlinks=true,
    linkcolor=BlueViolet,
    citecolor=Blue,
    urlcolor=MidnightBlue,
}

 \usepackage{cleveref}
\usepackage{stackengine}
\usepackage[normalem]{ulem}
\usepackage{slashed}
\usepackage{color}
\usepackage[dvipsnames]{xcolor}

\usepackage{bm}
\usepackage{braket}
\usepackage{slashed}
\usepackage{comment}
\definecolor{myblue}{rgb}{ 0.988, 0.078,0.458}

\newcommand{\eps}{\epsilon}
\newcommand{\chie}{\chi_e}
\newcommand{\chip}{\chi_p}
\newcommand{\Omg}{\hat{\Omega}\cdot {\hat p} + \vgbar}
\newcommand{\Omp}{\hat{\Omega}\cdot {\hat p} + \vpbar}

\newcommand{\odp}{ \hat \Omega\cdot \hat p}

\newcommand{\sx}{ \sigma_x}

\newcommand{\erf }{ \text{Erf }}
\newcommand{\U}{\chi_e}
\newcommand{\vpbar}{\bar{v}_p}
\newcommand{\vgbar}{\bar{v}_g}
\newcommand{\kbar}{\bar{k}}

\begin{document}  
\title{Testing Gravity with Realistic Gravitational Waveforms in Pulsar Timing Arrays}

\author{Wayne Hu}
\email{whu@kicp.uchicago.edu}
\affiliation{Kavli Institute for Cosmological Physics and Enrico Fermi Institute, The University of Chicago, Chicago, IL 60637, USA}
\author{Qiuyue Liang}
\email{qiuyue.liang@ipmu.jp}
\affiliation{Kavli Institute for the Physics and Mathematics of the Universe (WPI), University of Tokyo, Kashiwa 277-8583, Japan}
\author{Meng-Xiang Lin}
\email{mxlin@sas.upenn.edu}
\affiliation{Center for Particle Cosmology, Department of Physics and Astronomy, University of Pennsylvania, Philadelphia, Pennsylvania 19104, USA} 
\author{Mark Trodden} 
\email{trodden@upenn.edu}
\affiliation{Center for Particle Cosmology, Department of Physics and Astronomy, University of Pennsylvania, Philadelphia, Pennsylvania 19104, USA}

\date{\today}

\begin{abstract}
We consider the effects of relaxing the assumption that gravitational waves composing the stochastic gravitational wave background (SGWB) are uncorrelated between frequencies in analyses of the data from Pulsar Timing Arrays (PTAs). 
While uncorrelated monochromatic plane waves are often a good approximation, a background composed of unresolved astrophysical sources cannot be exactly uncorrelated since an infinite plane wave propagates no temporal signal.
We consider how relaxing this assumption allows us to extract potential information about modified dispersion relations and other fundamental physics questions, as both the group and phase velocity of waves become relevant. 
After developing the formalism we carry out simple Gaussian wavepacket examples and then consider more realistic waveforms, such as that from binary inspirals.  
When the frequency evolves only slowly across the PTA temporal baseline, the monochromatic assumption at an effective mean frequency remains a good approximation and we provide scaling relations that characterize its accuracy.
\end{abstract} 

\maketitle

\hrule
\tableofcontents
\hrule
\newpage

\section{Introduction} 
Pulsar Timing Arrays (PTAs) have opened up an entirely new window to observe any physics that modifies the propagation of signals from distant sources. This method allows for the detection of the stochastic gravitational wave background (SGWB), measurements of gravitational waves (GWs) from individual sources, constraints on the effects of fuzzy dark matter, and enables novel tests of fundamental physics \cite{NANOGrav:2023hvm}. Of particular interest is the recent 
$3\sigma$ evidence of a quadruple correlation in the signal, indicating the gravitational-wave nature of the timing residuals~\cite{NANOGrav:2023gor, EPTA:2023fyk, Reardon:2023gzh, Xu:2023wog}.

The primary quantity measured by PTAs is the two-point correlation function of timing delays in signals received from sources that are spatially separated on the sky. As we will review later, the angular dependence of this quantity, averaged over all sources, is generally referred to as the overlap reduction function (ORF). If we specialize to General Relativity (GR), then a particularly useful result is that information about the sources themselves can be well separated from the information contained in the ORF. 
In fact, both the SGWB and a single GW source can produce similar angular correlations in PTA surveys. 
For an isotropic SGWB, the ORF adopts a specific, well-known form, known as the Hellings-Downs curve \cite{Hellings:1983fr}. 
Interestingly, this same form is obtained for the PTA response to a single GW source when averaging over a statistically isotropic pulsar distribution \cite{Cornish:2013aba}.

This unique prediction of the Hellings-Downs curve when detecting gravitational waves using PTAs means that these systems provide a novel opportunity to test deviations from GR \cite{Liang:2023ary,Lee:2010cg,Gair:2015hra,Qin:2020hfy,Liang:2021bct,Bernardo:2022rif,Atkins:2024nvl,Bernardo:2023zna}. In modified gravity (MG), the ORF might display a different dependence on the angular separation of pulsar pairs. In the case of a modified dispersion relation for the graviton, only the phase velocity enters the relevant modification to the ORF if we assume the plane wave nature of the SGWB and no correlations between frequencies \cite{Liang:2023ary}. However, real signals cannot be monochromatic, since an infinite plane wave carries no information. Signals from real sources like binary inspirals arrive and propagate at their group velocity alongside the pulsar pulses, raising the question of how the group velocity and frequency correlations enter PTA timing for a background composed of astrophysical sources.  Furthermore, if new physics principles are at play, interesting new effects can arise away from the monochromatic limit. For example, if new physics leads to a modified dispersion relation for gravitational waves, then relaxing the assumption of a monochromatic source allows us to see how the effect of the group velocity can be modified compared to the predictions of GR.

With this in mind, in this paper we consider the question of whether it is justified to approximate the SGWB to be uncorrelated monochromatic plane waves, and investigate how the group velocity can change the ORF. This question depends crucially on the source of the gravitational waves. Both astrophysical sources such as supermassive black hole binaries (SMBHB) and various cosmological sources could give rise to a SGWB in the nanohertz band \cite{NANOGrav:2023hvm}. We would like to establish a formalism to study the effect of wave packets and further apply it to SMBHB sources to estimate how significant the effect could be in a realistic scenario.

The outline of the paper is as follows.
In the next section we review the formalism of pulsar timing residuals for the case of monochromatic GWs that superimpose incoherently. In \S~\ref{sec:Gaussian}, we then focus on the thought example in which the gravitational waves propagate as a Gaussian wavepacket. We use the width of the wavepacket as an expansion parameter to evaluate the PTA response signal at each order, and find that for a narrow wavepacket in the frequency domain, the net effect is still just a wavepacket superposition of the result obtained by assuming monochromatic plane waves at Earth and the pulsar, and the modification of Hellings-Downs curve of each frequency still only depends on the phase velocity. This can simply be thought of as an average over the modified Hellings-Downs angular dependence with the frequency spread of the signal. We then turn to the more realistic example of a supermassive black hole binary inspiral in \S~\Ref{sec:WKB}. We discuss the waveform with and without the distortion due to modified propagation effects and their impact on the ORF.   
We conclude with a discussion of these results in \S~\Ref{sec:conclude}.

Throughout the paper, we use the metric signature $(-,+,+,+)$, and set $\hbar = c = 1$.

\section{Monochromatic GWs}
\label{sec:mono}

We begin with a review of pulsar timing residuals for the case of monochromatic GWs that superimpose incoherently, following \cite{Liang:2023ary}. 
The direct observable in PTA is the pulse arrival time residual, relative to the start of the observation at $t=t_s$, 
\begin{equation}
\label{eq,residual}
    R(t) = \int_{t_s}^t  \frac{\nu_0-\nu(t')}{\nu_0} dt'\  \equiv \int_{t_s}^t z(t') dt'\ ,
\end{equation}
where the integrand is the fractional change in the pulse frequency $\nu$ from its unperturbed value $\nu_0$, and we will refer to it as the {\it redshift},\footnote{This is not the cosmological redshift, $z_s$, that we discuss in Eq.\ \eqref{eq,reshift}} $z$. 

The redshift is induced by metric perturbations that the pulse experiences while propagating along the geodesic from pulsar emission to reception. Writing the gravitational wave as $h_{ij}(\vec x, t)=e_{ij}^A(\hat\Omega) h^A(\vec x, t)$, 
with $ \hat\Omega$ denoting the GW propagation direction, the redshift can be written as
\begin{eqnarray}
\label{eq,redshiftpath}
 z &=& -\sum_{A= +,\times }\frac{\nu_0 \hat p^i \hat p^j e^A _{ij}}{2}  \int_{\lambda_e}^{\lambda_p} d\lambda \frac{\partial h^A (\lambda)}{\partial t}\ ,
\end{eqnarray}
where $\hat p$ is the unit vector denoting the pulsar direction, 
the summation over $A \in \{+,\times\}$ denotes the contribution from two polarization modes, and we will drop this sum in the following context when this does not lead to ambiguity. In the Cartesian basis of the GW wave $({\hat e}_1, {\hat e}_2, \hat \Omega)$, for example, the usual linear polarization states are given by
\begin{equation}
    e_{ij}^+( \hat\Omega) = \hat e_{1i} \hat e_{1 j}  - \hat e_{2 i} \hat e_{2 j} 
    \quad e_{ij}^\times( \hat\Omega) = \hat e_{1 i} \hat e_{2 j}  + \hat e_{2 i} \hat e_{1 j}  \ ,
\end{equation}
where the indices $i,j$ run over the transverse directions.
The signal traverses the distance $L$ from the pulsar to Earth in a time $t_e-t_p$, with
\begin{equation}
    t = \nu_0 (\lambda-\lambda_e) + t_e   \ , \ \ \hat\Omega \cdot \vec x =\hat\Omega \cdot \vec x_e - \nu_0 (\lambda-\lambda_e)  \hat\Omega\cdot \hat p\ ,
\end{equation} 
where subscripts $p,e$ denote that the relevant quantity is evaluated at the pulsar and Earth respectively, and $\hat p$ is defined by $\vec x_p - \vec x_e = L \hat p$.

For a monochromatic plane GW with momentum  $\vec k=k \hat\Omega$ and coordinates along the propagation direction of  $r=\hat\Omega\cdot \vec x$,
\begin{equation}
h^A(r,t) =  e^{i \phi( r, t)} h_0^A\ ,
\label{eq,mono}
\end{equation}
where the phase 
\begin{equation}
\label{eq,phase}
\phi({r, t}) =   k (r-r_0)  - \omega(k) (t-t_0)  \ ,
\end{equation}
is determined by the dispersion relation $\omega(k)$, and the reference point $r_0 = \hat \Omega \cdot \vec x_0$ and $t_0$, at which $\phi=0$.
Since we are here explicitly assuming a fixed monochromatic plane wave, we leave the $k$ and $\hat\Omega$ dependence of these quantities implicit and allow $z$ to be complex.
We will come back to the more general case where the GW is a superposition of multiple monochromatic modes and complex conjugation restores reality in Sec \ref{sec:Gaussian}.

Since Eq.~(\ref{eq,redshiftpath}) can be rewritten as the integral of a total derivative, it depends only on the end points in the form (see Eq.~(26) in Ref.~\cite{Liang:2023ary})
\begin{equation}
    z     = \frac{\hat p^i \hat p^j }{2} H(k)  [h_{ij}(\vec x_e, t_e)-h_{ij}(\vec x_p,t_p)]
      = \frac{\hat p^i \hat p^j e_{ij}^A (\hat\Omega)}{2}  H(k) h_0^A(e^{i\phi_e}-e^{i\phi_p}) \ ,
      \label{eq,monoz}
\end{equation}
where $\phi_e=\phi(r_e,t_e)$ and $\phi_p=\phi(r_p,t_p)$ are the GW phases when the pulse is received at Earth (at $t=t_e$) and is emitted at the pulsar (at $t=t_p$), respectively.
The GW dispersion relation only enters this expression through
\begin{equation}\label{eq,Hk}
H( k) = \frac{\omega(k)}{k\odp+\omega(k)}\ ,
\end{equation}
so only through the phase velocity $v_p(k) = \omega(k)/k$.
Between pulsars, the $\phi_p$ term generally contributes incoherently to the timing residuals, since $k L \gg 1$ and the phase difference becomes random unless
\begin{equation}
\Delta \phi = \phi_p -\phi_e = -i  L [k \odp + \omega(k)]
\label{eq,phasecoherence}
\end{equation}
is small ($|\Delta\phi|\ll 1$).  
The exceptional case is therefore when $k\odp+\omega(k) \rightarrow 0$, which can occur if $v_p<1$.   Even here the redshift remains finite, since $H(k)$ in Eq.~(\ref{eq,Hk}) has a cancelling pole.

Aside from this exceptional case, the relevant residual arrival time defined in Eq.~\eqref{eq,residual} for the plane wave case and an array of pulsars can be expressed as 
\begin{eqnarray}
\label{eq:R-mono}
    R(t,\hat p ) = \frac{\hat p^i \hat p^j e_{ij}^A}{2}  \frac{H( k)}{i\omega(k)} h_0^A  e^{i\phi_e}  \ ,
\end{eqnarray}
where $t=t_e$.  Here we keep the Earth term only and assume $t-t_s \gg 1/\omega$ so that the starting time of the observation drops out.

Following \cite{Roebber:2016jzl}, rather than averaging over random GW propagation directions for fixed pulsars, we can instead consider the pulsars to be randomly oriented with respect to a fixed GW when calculating residual correlations.
More explicitly, in spherical-polar coordinates with $\hat \Omega$ propagating toward the south pole,
we can express 
\begin{eqnarray}
    R(t,\theta,\phi) = \frac{1}{2} \frac{(1-\cos^2\theta) }{k  \cos\theta + \omega  } \left(\cos2\phi h_0^+  - \sin2\phi h^\times_0  \right) e^{i \phi_e}\, ,
\end{eqnarray}
where we have taken
\begin{equation}
\hat p =(\sin\theta\sin\phi, \sin\theta\cos\phi,\cos\theta)\, ,\text{and, } \ \hat\Omega = (0,0,1)\ .
\end{equation}
We then decompose this quantity into  spherical harmonics,
\begin{equation}
    a_{\ell m}(t) = \int d^2\hat p~ R(t,\theta,\phi) Y_{\ell m}^*(\theta,\phi) = \frac{1}{\omega } c_\ell (v_p(t)) \frac{\pi }{2} \sqrt{\frac{2 \ell +1}{4 \pi} \frac{(\ell -2) !}{(\ell +2) !}} \left( h_0^{+ } \pm i  h_0^{\times}\right) , \quad m= \pm 2 \ .
\end{equation}
Here the integration $\int d^2\hat p$ is over pulsar positions, and $c_\ell$ are the coefficients 
\begin{eqnarray}
\label{eq,cell}
    c_\ell (v_p) = \int_{-1}^1 d\cos\theta \frac{v_p}{v_p+ \cos\theta}
    (1-\cos\theta^2)^2 \frac{d^2}{d\cos\theta^2} P_\ell(\cos\theta)\ ,
\end{eqnarray}
which take the form\footnote{See Eq.~(16) in \cite{Liang:2024mex} for further discussion and the case of $v_p<1$. See also \cite{Domenech:2024pow} for a finite distance effect discussion.}
\begin{eqnarray}
   c_{\ell}= -2 v_p(1+\ell)\left(\left((2+\ell) v_p^2-\ell\right) Q_{\ell}(-v_p)+2 v_p Q_{\ell+1}(-v_p)\right)\ ,
   & \text { if } v_p \ge 1\ ,
\end{eqnarray}
where $Q_\ell(z) =\,$LegendreQ$[\ell,0,3,z]$ is the Legendre $Q$ function of the third type with $m=0$.  

For small deviations, $v_p = 1+ \varepsilon >1$, we can expand $c_\ell$ with respect to $\varepsilon$,  
\begin{eqnarray}\label{eq:cl-expansion}
   (-1)^\ell c_\ell &= & 4  -2  (\ell^2+\ell-4)\varepsilon -\frac{1}{2} \frac{(\ell+2)!}{(\ell -2)!}  \left( \log\frac{\varepsilon}{2} +2 \psi(\ell)+2\gamma_E \right)\varepsilon^2  \nonumber\\
&&-  \frac{1}{4} \left(12-\ell(3\ell-1)\right)(\ell+2)(\ell-1) \varepsilon^2 + \mathcal{O}(\ell^6\varepsilon^3)\ ,
\end{eqnarray}
where $\gamma_E$ is Euler's constant, and $\psi $ is the digamma function defined as 
\begin{eqnarray}
    \psi (z)  =\frac{\Gamma'(z)}{\Gamma(z)}\ .
\end{eqnarray}
In practice, we want to extract the angular correlation between two pulsars, assuming statistical isotropy of pulsar distribution:
\begin{equation}\label{eq:RR-mono}
    \langle R(t_1,\hat p_1) R^*(t_2,\hat p_2) \rangle \propto \Gamma(\xi) e^{i \omega (t_1-t_2)} \ ,
\end{equation}
where $\xi=\cos^{-1}(\hat p_1\cdot\hat p_2)$ is the angle between the two pulsars. The function $\Gamma(\xi)$ is called the {\it overlap reduction function} (ORF) and is conventionally normalized to $\Gamma(0)=1/2$.
Using this normalization, we then have 
\begin{equation}
\label{eq,angularcorr}
  \Gamma_{\rm m}(\xi) 
  \equiv \frac{1}{2} \frac{ \sum_{\ell=2}^\infty \frac{2\ell+1}{4\pi} C_\ell P_\ell(\cos\xi)} {\sum_{\ell=2}^\infty \frac{2\ell+1}{4\pi} C_\ell} \ ,
\end{equation}  
with the total power spectrum 
\begin{eqnarray}
\label{eq,Cl}
    C_\ell=\frac{1}{2 \ell+1} \sum_m a_{\ell m}^* a_{\ell m} = \frac{\pi}{8} \frac{|c_\ell(v_p )|^2}{\omega^2} \frac{(\ell-2)!}{(\ell+2)!}\sum_{A= +,\times }|h_0^A|^2  \ .
\end{eqnarray}
In the particular case of General Relativity, where $v_p(k)=1$, the overlap reduction function is given by the {\it Hellings-Downs} form \cite{Hellings:1983fr}
\begin{eqnarray}
    \Gamma_{\rm GR}(\xi) &=& \frac{3}{2} \sum_{\ell=2}^\infty \frac{2(2\ell+1)}{(\ell+2)(\ell+1)\ell(\ell-1)}P_\ell(\cos\xi) 
    \, \nonumber\\
    &=& \frac{1}{2} \left[ 1+\frac{3}{2} (1-\cos\xi)
    \left(\ln \frac{1-\cos\xi}{2}-\frac{1}{6} \right)\right] \,.
\end{eqnarray}

\begin{figure}   
\centering
\includegraphics[width=0.6\columnwidth]{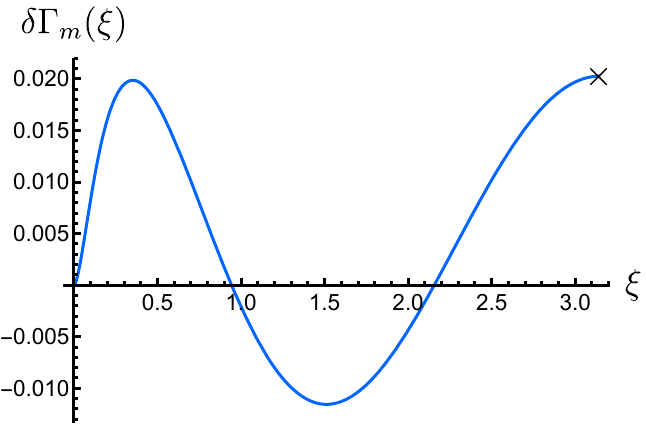}
    \caption{
    The difference between the overlap reduction function in GR and that in a modified gravity theory with a higher phase velocity $v_p-1=\varepsilon=0.01$. The $\times$ symbol at the antipodal point $\xi = \pi$ denotes the deviation $\delta\Gamma_m(\pi)\approx 0.02$ as predicted in Eq.\ \eqref{eq,deltaGamma}, which then is a good proxy for the overall amplitude of the deviations.  
    }
    \label{fig:deltaGamma}
\end{figure}

In the limit of a small positive deviation of the phase velocity from the speed of light, $v_p=1+\varepsilon$, the
leading order distortions to the Hellings-Downs curve scale as $\varepsilon$ but with a logarithmic running of the coefficient due to the summation over $\ell$ in Eq.~(\ref{eq,angularcorr}).  
In  Fig.\ \ref{fig:deltaGamma}, we show an example for $\varepsilon=0.01$.
Note that aside from zero crossings in $\Gamma_m$, the deviations are the same order of magnitude for all $\xi$, so a simple characterization of their amplitude is the deviation at $\xi=\pi$, for which 
\begin{eqnarray}
\label{eq,deltaGamma}
    \delta \Gamma_m(\pi) \equiv \Gamma_m(\pi)- \Gamma_{\rm GR}(\pi)\approx   - (0.6+0.57 \log\varepsilon) \varepsilon\ ,
\end{eqnarray}
is a good approximation if $\varepsilon < 0.07$ as can be seen in Fig.~\ref{fig:deltaGamma} (cross).

\begin{figure}   
\centering
\includegraphics[width=\columnwidth]{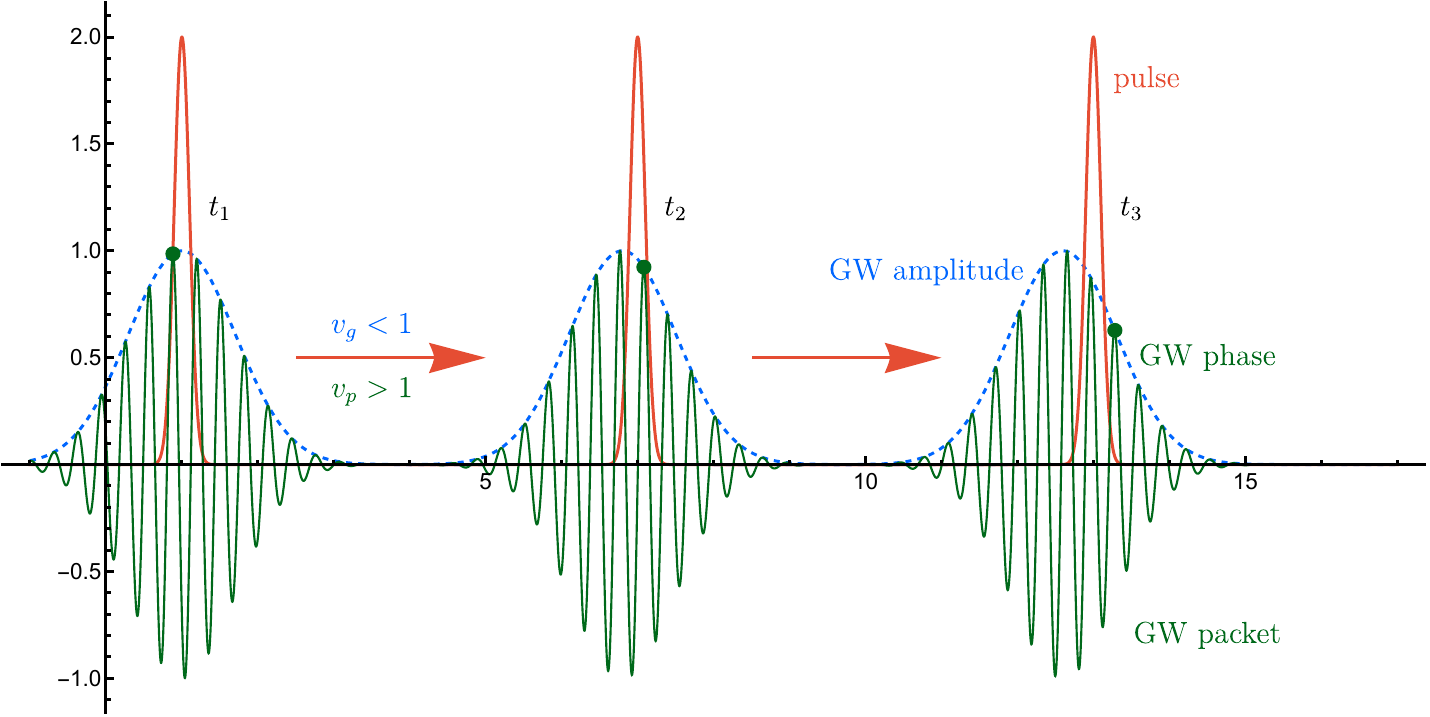}
    \caption{
    A sketch to illustrate  GW packet propagation (green lines)  along the pulsar pulse path.
    The red lines depict the pulse propagating at the speed of light. The blue lines indicate the envelope of the wavepacket and track the peak amplitude which propagates at a group velocity $v_g<1$ and lags the pulse. The green dots indicate a specific initial phase value ($\phi(r,t_0)=0 $) that propagates at a phase velocity $v_p>1$ and leads the pulse.
    }
    \label{fig:evolution}
\end{figure}

More generally, beyond GR the overlap reduction function becomes dependent on the GW momentum and dispersion relation. If we assume a monochromatic plane wave, modifications from the dispersion relation, including those to the overlap reduction function $\Gamma(\xi)$, can only enter through the phase velocity, $v_p(k)$. However, with a nontrivial dispersion relation, the above calculation can only be an approximation to the effect of waves generated by physical sources. Gravitational waves from a real source must propagate as a wavepacket at the group velocity $v_g = \partial \omega/\partial k$, and this fact should somehow enter into the observable redshift.  For example, when  $v_g < 1$, a GW signal peaking at pulsar emission may have no contribution to the redshift at Earth reception if it has yet to arrive at Earth; $h_{ij}(\vec x_e,t_e)\approx 0$, and it instead provides a source along the pulse propagation path $\partial h_{ij}/\partial t \ne 0$ for some $\lambda_e < \lambda < \lambda_p$ (see also Fig.~\ref{fig:evolution}). 

On the other hand, the locality of the monochromatic redshift expression in  Eq.~(\ref{eq,monoz}) suggests that the redshift for any wavepacket can be obtained from the coherent superposition of its plane wave components at pulsar emission and Earth reception alone, without further consideration of the propagation of the GW.
In the next section we shall use the example of a Gaussian wavepacket to reconcile these points of view before proceeding to examine the impact of GW propagation and group velocity on a realistic binary inspiral signal.

\section{Gaussian wavepacket}\label{sec:Gaussian}

We begin with a simple generalization of the monochromatic plane wave in Eq.\eqref{eq,mono} to build intuition about pulsar timing residuals with real signals.   Consider an initial wavepacket at $t=t_0$ where the amplitude of the originally monochromatic wave at $k=\kbar$ is modulated by a spatial Gaussian of width $\sigma_x$ 
\begin{equation}
h^A(r, t_0) = e^{-\frac{(r-r_0)^2}{2\sx^2}}  e^{i \kbar (r-r_0)}h_0^A  \ .
\label{eq,Gaussinit}
\end{equation}
There are two ways to think about the effect of such a wavepacket on pulsar timing. One is to note that the center of the wavepacket propagates at the group velocity\footnote{This can be obtained from the stationary phase point, see Eq.~(\ref{eq:SPP}).} $\vgbar = \partial \omega/\partial k|_{\kbar}$.
Therefore, when performing the integral along the pulse path to obtain the redshift Eq.~(\ref{eq,redshiftpath}), the Gaussian modulation changes the GW contribution between the pulsar and Earth according to the spacetime position of the wavepacket. 

On the other hand, we can decompose the wavepacket into its monochromatic components via a Fourier transform 
\begin{equation}
h^A(k, t_0)= \int d r\,  e^{-i k r} h^A(r, t_0) = 
{\sqrt{2\pi}}{\sigma_x}  e^{-ik r_0} e^{- (k-\kbar)^2\sigma_x^2/2} h_0^A \ ,
\label{eq,hAk}
\end{equation}
which implies that there is a Gaussian spread of wavenumbers around $\kbar$. Note that since we have fixed the GW direction, we apply a one dimensional Fourier transform instead of the 3D version. Notice also that we allow $h^A(r,t)$ and hence $z(t)$ to be complex for notational simplicity. Our convention is that to obtain a real signal we add the complex conjugate or, more generally when composing a full wavepacket, we compute for $k>0$ only and implicitly assume that  $h^A(-k)=h^{A*}(k)$. Hence
\begin{eqnarray}
    h^A(r,t) &=& \int_{-\infty}^\infty \frac{dk}{2\pi} e^{i kr} h^A(k,t )=  2\int_0^\infty \frac{dk}{2\pi}\Re\left[e^{i kr}h^A(k,t ) \right] \ ,
    \label{eq:hFourier-real}
\end{eqnarray}
where $k = |\vec k|$, and $r =\hat\Omega \cdot \vec x$.

Each of these components should affect the observable redshift through the monochromatic expression (\ref{eq,monoz}) which can then be resuperimposed at the pulsar and Earth endpoints.   
We now use this Gaussian example to illustrate why these pictures are equivalent and how the group velocity enters into each.

\subsection{The propagation approach}
In the stationary phase approximation, the Gaussian wavepacket defined at $t_0$ in Eq.~(\ref{eq,Gaussinit}) propagates according to the group velocity via
\begin{equation}\label{eq,hA}
h^A(r,t) = e^{i \bar\phi(r,t)-\chi^2(r,t)/2}h_0^A\ ,
\end{equation}
where 
\begin{equation}
\bar\phi(r,t) = \bar k (r-r_0) - \omega(\bar k) (t-t_0)
\end{equation}
propagates the initial phase at the phase velocity and $\chi(r,t)$ tracks the dimensionless distance from the center of the wavepacket in units of $\sigma_x$,
\begin{equation}
\label{eq,dimensionlessdis}
\chi(r,t) =  \frac{ (r-r_0) - \vgbar (t-t_0) }{\sigma_x} \ .
\end{equation}
Here $\vgbar = \partial\omega/\partial k|_{k=\kbar}$ is the group velocity at central momentum. 

In Fig.~\ref{fig:evolution} we illustrate the propagation of this wavepacket and compare it to the pulsar pulse which travels at the speed of light.  After propagation, the center of the wavepacket, where the GW amplitude is the highest, lags the pulse if $\vgbar<1$ whereas a given initial phase value leads the pulse if $\vpbar>1$.
Moreover, the phase at the center of the wavepacket evolves with time.  Eventually, the pulse will lead the GW packet by such a large spatial separation that the GW amplitude at the pulse and at its reception on Earth is negligible.

We can explicitly evaluate the redshift in Eq.~(\ref{eq,redshiftpath}) using our Gaussian wavepacket Eq.~(\ref{eq,hA}) as, 
\begin{eqnarray}
\label{eq,redshift}
 z &=& -\frac{\nu_0 \hat p^i \hat p^j e^A _{ij}}{2}  \int_{\lambda_e}^{\lambda_p} d\lambda \frac{\partial h^A (\lambda)}{\partial t}\nonumber\\
 &=& -\frac{\nu_0\hat p^i \hat p^j e^A _{ij}}{2}  \int_{\lambda_e}^{\lambda_p} d\lambda  \left[-i \omega(\kbar) + \frac{\vgbar}{\sigma_x^2}\left((r-r_0)
 -\vgbar(t-t_0) \right)\right] h^A(\lambda) \ .
\end{eqnarray}
Performing the integration directly gives, 

\begin{eqnarray}
\label{eq,redshift2}
    z&=& -\frac{  \hat p^i \hat p^j e^A _{ij}}{2} h_0^A  e^{i \bar\phi_e  -\frac{\U^2}{2}  } \nonumber\\
   & \times&\left(\frac{B}{2D} \left(1- e^{-CL - DL^2}\right) + \frac{BC+2AD}{4D^{3/2}} \sqrt{\pi}e^{\frac{C^2}{4D}} \left(\erf\frac{C}{2\sqrt{D}} -\erf\frac{C+2DL }{2\sqrt{D}}   \right) \right)\ ,
\end{eqnarray} 
with
\begin{eqnarray}
   && A = - i \vpbar \kbar \left(1 + \frac{i}{\sx \kbar} \frac{\vgbar}{\vpbar} \U \right)\ , \ \ \ \   B= - \frac{\vgbar}{\sx^2} (\odp+\vgbar) \ , \nonumber\\
  && C= - i \kbar \left(\odp+ \vpbar\right)\left(1+ \frac{i}{\sx \kbar} \U \frac{\odp+\vgbar}{\odp+\vpbar}\right)\ , \ \ \ D = \frac{1}{2\sx^2} (\odp+\vgbar)^2 \ .
\end{eqnarray} 
Here
\begin{equation}
\chie  \equiv \chi(r_e-r_0  ,t_e-t_0)\ ,
\end{equation} 
gives the dimensionless distance between the Earth and the center of the wavepacket at the time of reception, and, for later convenience, 
\begin{equation}
\chip  \equiv \chi(r_p-r_0  ,t_p-t_0)\ ,
\end{equation} 
gives the dimensionless distance between the pulsar and the center of the wavepacket at the time of pulsar emission.
Notice that the redshift takes the form of the difference between an $L$-independent Earth term and an $L$-dependent pulsar term.

Now, in the limit $\sx \to \infty$, the wavepacket reduces to a monochromatic wave and the redshift is the same as Eq.~(\ref{eq,monoz}), 
\begin{eqnarray}
\label{eq,limitphase}
   \lim_{\sigma_x\rightarrow \infty}  z = \frac{ \hat p^i \hat p^j  e^A _{ij}}{2} H(k) \left[
    e^{i  \bar\phi_e}  
-
    e^{i  \bar\phi_p} \right]h_0^A \ ,
\end{eqnarray}
where the phases at Earth and at the pulsar are respectively given by $\bar \phi_e=\bar\phi(r_e,t_e)$, and $\bar  \phi_p=\phi(r_p,t_p) = \bar\phi_e+  \kbar L (\Omp)$. 
Away from this limit, the spatial modulation broadens the range of frequencies in the wavepacket. If we assume that this frequency spread is small, 
\begin{equation}
\eps \equiv \frac{1}{\kbar \sigma_x} \ll 1\ ,
\end{equation}
such that  $| \eps \chie| \ll 1$ and $| \eps \chip| \ll 1$, we can expand Eq.\ \eqref{eq,redshift2} with respect to $\eps$ as
\begin{eqnarray}
    z = \frac{ \hat p^i \hat p^j e^A _{ij}}{2} h_0^A  e^{i  \bar\phi_e -\U^2/2}  
    \sum_{n=0} B_{n,e} \frac{\epsilon^n k^n }{n!} \frac{\partial^n  H( k)}{\partial   k^n}\bigg|_{k=\kbar} -\{e\rightarrow p\} \ , 
    \label{eq,zpropagation}
\end{eqnarray}
where the first term is the contribution from the Earth term, and $\{e\rightarrow p\}$ corresponds to the symmetric pulsar term. 
The coefficients $B_{n,e}$ are given by
$B_{0,e}=1$, $B_{1,e}=i \U $,  
$B_{2,e}=(1-\U^2) $, 
$B_{3,e}=i \U(3-\U^2) $, 
$B_{4,e}=(3-6 \U^2 + \U^4 )$, etc.\ and likewise for $\{e\rightarrow p\}$.
See the next subsection for a more general expression for $B_{n,e}$.
Notice that the conditions $| \eps \chie| \ll 1$ and $| \eps \chip| \ll 1$ allow $\chie,\chip>1$; i.e.\ for sufficiently small frequency spread $\eps$, these conditions allow for the reception at Earth and the emission at the pulsar both to be in the tails of the Gaussian, and the wavepacket to lie in-between.   Nevertheless, the redshift can still be divided into terms that are local at these two events.

The physical interpretation of this form of the redshift is relatively simple. The $n=0$ term is associated with the change in the amplitude and phase of the GW between pulsar emission and Earth reception through their respective phase, $\bar \phi(r,t)$, and distance, $\chi(r,t)$, factors.
The $n=1$ term represents the leading order Taylor correction to $H( \kbar)$ to account for the finite spread of frequencies in the wavepacket.   This term then carries the dependence on the group velocity, as we can see from
\begin{equation}
  \frac{\partial \ln H( k ) }{\partial \ln k}  \Big|_{k=\kbar} =  \frac{\vgbar}{\vpbar}- \frac{\Omg}{\Omp} \ ,
\end{equation}
and vanishes if the phase and group velocities are equal. In this sense the redshift is just the wavepacket-weighted average of the monochromatic result. We show next that this can be obtained directly from superimposing its monochromatic components at pulsar emission and at Earth reception.  Notice also that the limit implicitly assumes that $H(k)$ is smooth and the Taylor expansion valid, which excludes the singular case discussed around Eq.~(\ref{eq,phasecoherence}),  where
$k\hat\Omega\cdot \hat p + \omega(k) =0$, as we shall also see more directly next.

\subsection{The superposition approach}\label{sec:WKB}
Given the monochromatic components of the wavepacket from its Fourier transform Eq.~(\ref{eq,hAk}), we can alternatively superimpose the monochromatic redshifts of Eq.~(\ref{eq,monoz}) with the Fourier weights and phases at Earth reception and pulsar emission separately, via
\begin{equation}
z =\frac{\hat p^i \hat p^j e_{ij}^A}{2} (I_e-I_p) \ ,
\label{eq,zsuper}
\end{equation}
where
\begin{equation}
I_e =
 \int \frac{dk}{2\pi}
H( k)  h^A(r_e,t_e;k) \ ,
\label{eq,superposition}
\end{equation}
and similarly for $I_p$ with $e\rightarrow p$.
For each $k$, we propagate the initial Fourier amplitude independently 
\begin{equation}\label{eq:hA-superposition}
h^A(r,t;k)  =  e^{i k r- i \omega(t-t_0) }h^A(k,t_0)
= e^{i\phi(r,t;k)}{\sqrt{2\pi}}{\sigma_x} e^{- (k-\kbar)^2\sigma_x^2/2} h_0^A \ ,
\end{equation} 
with a phase factor
\begin{equation}
\phi(r,t;k) = k (r-
r_0) - \omega (t-t_0) \ .
\end{equation}
In particular, we employ the phases at pulsar emission and Earth reception
\begin{equation}
\varphi_e(k) = \phi(r_e,t_e;k)\, ,\qquad
\varphi_p(k) = \phi(r_p,t_p;k) \ ,
\end{equation}
and note that for comparison with the propagation approach Eq.\eqref{eq,limitphase}, $\varphi_e(\kbar)=\bar \phi_e$, $\varphi_p(\kbar)=\bar \phi_p$.

Let us explicitly evaluate the Earth term by again assuming that the spread of frequencies in the Gaussian wavepacket is small compared to the $k$-dependence of other quantities in Eq.~(\ref{eq,superposition}).
By expanding
\begin{equation}
e^{i\varphi_e(k)} \approx e^{i[\bar\phi_e  + \varphi_e'(\kbar)(k-\kbar)]} \left[1+\frac{i}{2} \varphi_e''(\kbar) (k-\kbar)^2 + \ldots \right]
\end{equation} 
and using
\begin{equation}
\varphi_e'(\kbar) = \frac{d\varphi_e}{dk}\bigg|_{\kbar}= \chi_e \sigma_x \ ,
\end{equation}
we obtain to second order in the Taylor expansions of $H(k)$ and the phase
\begin{equation}
\label{eq:I-WKB}
\frac{I_e}{ h_0^A e^{i \bar\phi_e } e^{-\chi_e^2/2} }
= H(\kbar)+ i H'(\kbar)\chi_e \sigma_x^{-1}
+\frac{1}{2} (1-\chi_e^2) (H''(\kbar)+ i\varphi_e''(\kbar) H(\kbar))\sigma_x^{-2} +\ldots
\end{equation}
and likewise for the pulsar term with $e \rightarrow p$.   
This has the same coefficients $B_{n,e}$ as the propagation equation Eq.~(\ref{eq,zpropagation}) aside from the $\varphi''$ terms. 
More generally, we have
\begin{eqnarray}\label{eq:Bn}
    B_{n,e} =  i^n \chi_e 2^{n-1/2}    U\left(\frac{1-n}{2}, \frac{3}{2}, \frac{\chi_e^2}{2}\right)\ ,
\end{eqnarray}
where $U(a,b,z)$ is the confluent hypergeometric function.
We provide the detailed derivation in Appendix \ref{app:Bn}. 
In the stationary phase approximation these terms represent the group velocity dispersion and are higher order corrections. Group velocity dispersion broadens the initial wavepacket width $\sigma_x$  during propagation after a time
\begin{equation}
t-t_0 \gtrsim \sigma_x^2/ (\partial^2 \omega/\partial k^2)\ .
\label{eq,groupvelocitydispersion}
\end{equation}
For the Gaussian wavepacket case, we can simply renormalize $\sigma_x$ to its value near the Earth-pulsar system in the Galaxy and ignore the broadening effect from gravitational wave propagation from the source to the Galaxy \cite{Ezquiaga:2021ler}.  We shall return to this issue for the inspiral GW in \S~\ref{sec:realwaveform}, where formally the spatial wavepacket is very broad but only a small portion of the total can affect pulsar timing across the observation baseline.

Notice also that we have explicitly assumed that $H(k)$ itself can be Taylor expanded and this is not true if $k\hat\Omega\cdot \hat p + \omega(k) =0$ where the packet has support (see Eq.~(\ref{eq,phasecoherence})).  In this case the integral over $k$ in $I_e-I_p$ for Eq.~(\ref{eq,zsuper}) must be carried out jointly and $H(k) [e^{i\varphi_e}-e^{i\varphi_p}]$ remains smooth.

The group velocity emerges from the monochromatic expression via the derivatives of the dispersion relation in the Taylor expansion, as well as the characterization of the amplitude of the gravitational wave at Earth and pulsar, and matches the explicit appearance in the wavepacket propagation approach.   We again see that the redshift associated with the total wavepacket is simply the wavepacket-weighted average over the monochromatic expressions.

\section{Inspiral Waveform}\label{sec:realwaveform}
In this section, we analyze the PTA signal of a binary inspiral GW using the same wavepacket formalism as with the Gaussian wavepacket of the previous section. As in the Gaussian case, the induced pulsar redshift can be thought of as the superposition of the redshifts of the individual monochromatic components of the wavepacket that then modifies the overlap reduction function.

On the other hand, the GW signature of a binary inspiral has features that differ from the Gaussian wavepacket example in a number of important ways.  First, as discussed in \S~\ref{sec:temporal}, the waveform is usually characterized in the temporal domain rather than spatial domain, and its form for the inspiral case is given in \S~\ref{sec:nearemission}.  The temporal frequency content of the full wavepacket can be very broad, as discussed in \S~\ref{sec:inspiral-redshift} as the emission evolves from the inspiral phase to the coalescence one -- so broad that group velocity dispersion can make the waveform distort dramatically when propagating over cosmological distances.
Nonetheless, at any given epoch it is mostly composed of a single frequency for the dominant quadrupole mode: that of the stationary phase point.   
As we show in \S~\ref{sec:example},
this single frequency per observation time can in principle evolve across the pulsar timing baseline and generate correlations between different frequencies in the signal, each with their own distortion to the Hellings-Downs curve. Hence the arrival times of different frequency components and their determination through the stationary phase approximation become the critical quantities of interest.

\subsection{Temporal vs.~spatial domain}
\label{sec:temporal}

The first difference between the Gaussian wavepacket example and the case of a binary inspiral is that the inspiral wavepacket is usually described in the time domain rather than the spatial domain or analogously by its temporal frequency ($\omega$) content at a fixed position rather than its spatial frequency ($k$) at a fixed time.  To connect these two descriptions,
we can write (see Eq.~\eqref{eq:hFourier-real})
\begin{eqnarray}
    h^A(r ,t) = \int_{-\infty}^\infty \frac{d\omega}{2\pi}  \frac{\partial k}{ \partial  \omega}   e^{i(k r -\omega t)}h^A(k(\omega), t=0 ) \ ,
\end{eqnarray}
and then take the inverse temporal Fourier transform at some fiducial $r=0$ point 
\begin{eqnarray}
    h^A(r=0 ,\omega) &=& \int_{-\infty}^{\infty} dt e^{i\omega t} h^A(r=0 ,t)=   \frac{\partial k(\omega)}{\partial \omega}  h^A(k(\omega),t=0) \nonumber \\
    &\equiv & 
    \tilde h(f)
    \ ,
\end{eqnarray}
where in the last line, we have given the temporal Fourier transform a distinct label $\tilde h(f)$, with the observationally received frequency $f=\omega/2\pi$ to avoid notational confusion below when relating it to the spatial Fourier transform and the propagation angular frequency $\omega$.\footnote{We do not write the polarization dependence explicitly in $\tilde h(f)$ or in the time domain waveform, $h(t)$ for notational simplicity.}

In this relation, the group velocity $\partial \omega/\partial k$ acts as the Jacobian transformation between the two spaces, and we have traded spatial evaluation at a fixed time $h^A(k,t=0)$ for temporal evaluation at a fixed position $h^A(r=0,\omega)=\tilde h(f)$. If there is no screening effect in the PTA system, and the gravitational theory is described by the modified gravity theory both at the pulsar and at Earth, then this Jacobian would also map the spatial power spectrum of the gravitational wave background at emission to the temporal power spectrum at reception. 

In practice, we use this transformation to take a time domain waveform and extract its temporal frequency representation, $\tilde h(f)$, and infer the spatial frequency content, $h^A(k,t=0)$, for the evaluation of the PTA overlap reduction function.  
For example, for the relevant Earth term we choose coordinates where $r_e=0$ and $t=t_e$ to obtain
\begin{align}
I_e(t) ={}&\int_{-\infty}^{\infty} \frac{dk}{2\pi} \, h^A(r=0,t_e ; k ) H(k) = 2\int_0^\infty df\, \Re[  \tilde h(f)  e^{-i 2\pi f t }  H(k(f)) ] \, .
\end{align}
Furthermore, to the extent that the waveform obeys the SPA, each frequency arrives at a given time and this integral using the SPA just returns the evaluation of $H(k)$ at the arrival frequency $f(t)$ mapped back to the spatial wavenumber $k(t)$ times the original time domain waveform:
\begin{equation}
I_e(t) = h^A(r=0,t) H(k(t))\, ,
\end{equation}
where $h^A(r=0,t)$ carries the amplitude and phase of the GW at the Earth,
as one would expect. The same prescription would apply to $I_p(t)$ but now the gravitational waveform is described by the amplitude and phase at the emission time of the pulsar. Since the pulsar term does not contribute coherently between pulsars except in the special case of Eq.~(\ref{eq,phasecoherence}), from this point forward we focus exclusively on the Earth term and implicitly take Earth centric coordinates $r_e=0$ when evaluating $\tilde h(f)$ and its time domain Fourier pair $h(t)$.

Notice that the PTA timing correlation between two different times as in Eq.~(\ref{eq:RR-mono}) now corresponds to two arrival frequencies $f(t_1)$ and $f(t_2)$ that embed the dependence on the group velocity.  As long as this evolution is slow compared with the millisecond pulsar clocks themselves, we can consider the residuals at each epoch to be that of a monochromatic plane wave.   The remaining modifications of the overlap reduction function will then 
be characterized by the evolution of the frequency across the observational baseline, which we set to reflect current constraints, $T_{\rm obs} \sim 15$yrs,
\begin{equation}
\Delta \ln f= \frac{d\ln f}{dt} T_{\rm obs} \, .
\end{equation}
Below, we will occasionally use the shorthand convention $H(f) \equiv H(k(f))$ and
${d\ln H}/{d \ln f}$ where doing so leads to no confusion.
Given this dependence, we next consider the frequency evolution and validity of the SPA for binary inspiral waveforms.

\subsection{Time domain and near emission waveform}
\label{sec:nearemission}

We adopt the binary black hole inspiral waveform predicted by GR \cite{Cutler:1994ys} and assume that any modified gravity effect is screened in the emission region. 
Under GR or before modified propagation effects become important, the received time domain waveform can be described as 
\begin{eqnarray}\label{eq:TD-ht}
    h(t) = 2\sqrt{\frac{6}{5}}\frac{G \mathcal{M }\pi^{2/3} Q }{ D }   \left(\frac{t_c - t}{5 G \mathcal{M} }\right)^{-\frac{1}{4}}  e^{i\phi(t)} \ ,   
\end{eqnarray}
where $Q$ contains the geometric information about the binary system, such as the inclination angle and the polarization angle, $t_c$ is arrival time of the signal from the coalescence of the system, 
$G$ is the Newtonian gravitational constant, $\mathcal{M} = (M_1 M_2)^{3/5}/(M_1+M_2)^{1/5}$ is the chirp mass for a binary black hole system with masses $M_1$ and $M_2$, $D$ is the distance to the source, and $\phi$ is the time-dependent phase of the gravitational wave,  
\begin{equation}
\label{eq,timedepphase}
    \phi(t) = -2\left( \frac{t_c-t}{5 G\mathcal{M}}\right)^{\frac{5}{8} }+\phi_c \ ,
\end{equation}
with $\phi_c=\phi(t_c)$. This inspiral waveform assumes $t\ll t_c$ and $f<f_c = (6^{3/2} \pi G \mathcal{M})^{-1}$, and we shall see below how to remove higher frequencies by apodization.

With this form, the frequency 
of the GW evolves according to
\begin{equation}
    \frac{df}{dt} = \frac{96}{5}\pi^{8/3}\left(G \mathcal{M}\right)^{5/3}f^{11/3}\ .
\end{equation}
Notice that the equations above are invariant under the transformation
\begin{equation}
\label{eq,reshift}
    [\mathcal{M},f,t,D]\rightarrow[\mathcal{M}(1+z_s), f/(1+z_s), t(1+z_s), D(1+z_s)] \ ,
\end{equation} 
where $z_s$ is the cosmological redshift of the source. Therefore, from the observed waveform we can only extract the redshifted chirp mass $\mathcal{M}=\mathcal{M}_{\rm true}(1+z_s)$ and the comoving distance $D=\int_0^{z_s} dz_s'/H(z_s')$ where $H(z)$  here is the Hubble parameter and should not be confused with the function $H(k)$ that modifies the pulse redshift $z$. From this point forward, the binary masses and distances in this paper will refer to these quantities.

For large distances, the waveform will distort and change its phase due to modified propagation.  Given the dispersion relation, this is simple to account for in the frequency domain. 
To move between the time and frequency domains requires the stationary phase approximation, which we next show is a good approximation in the case of the inspiral waveform for $f\ll f_c$.

\subsection{Frequency domain and stationary phase approximation }\label{sec:inspiral-redshift}

\begin{figure}   
\centering
\includegraphics[width=0.7\textwidth]{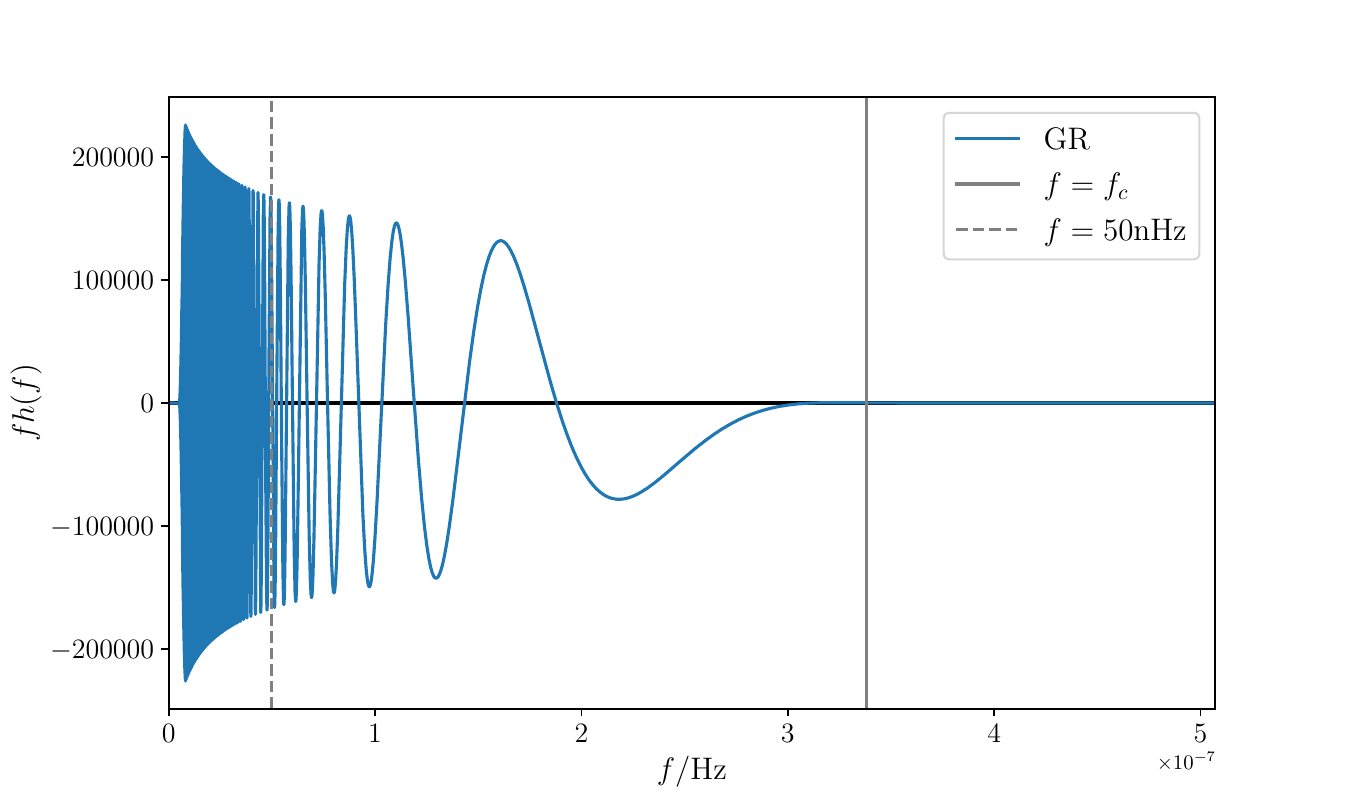}
    \caption{The frequency domain waveform for the binary inspiral  following Eq.~\eqref{eq:FD-hf} with a redshifted chirp mass $\mathcal{M}=1.3\times 10^{10}M_\odot$. We apply apodization at both the high and low-frequency ends (see text) and highlight the positions of the coalescence frequency $f_c$ as well as the typical PTA relevant frequency $f=50\,$nHz.}
    \label{fig:FD-hf}
\end{figure}

\begin{figure}   
\centering
\includegraphics[width=0.7\textwidth]{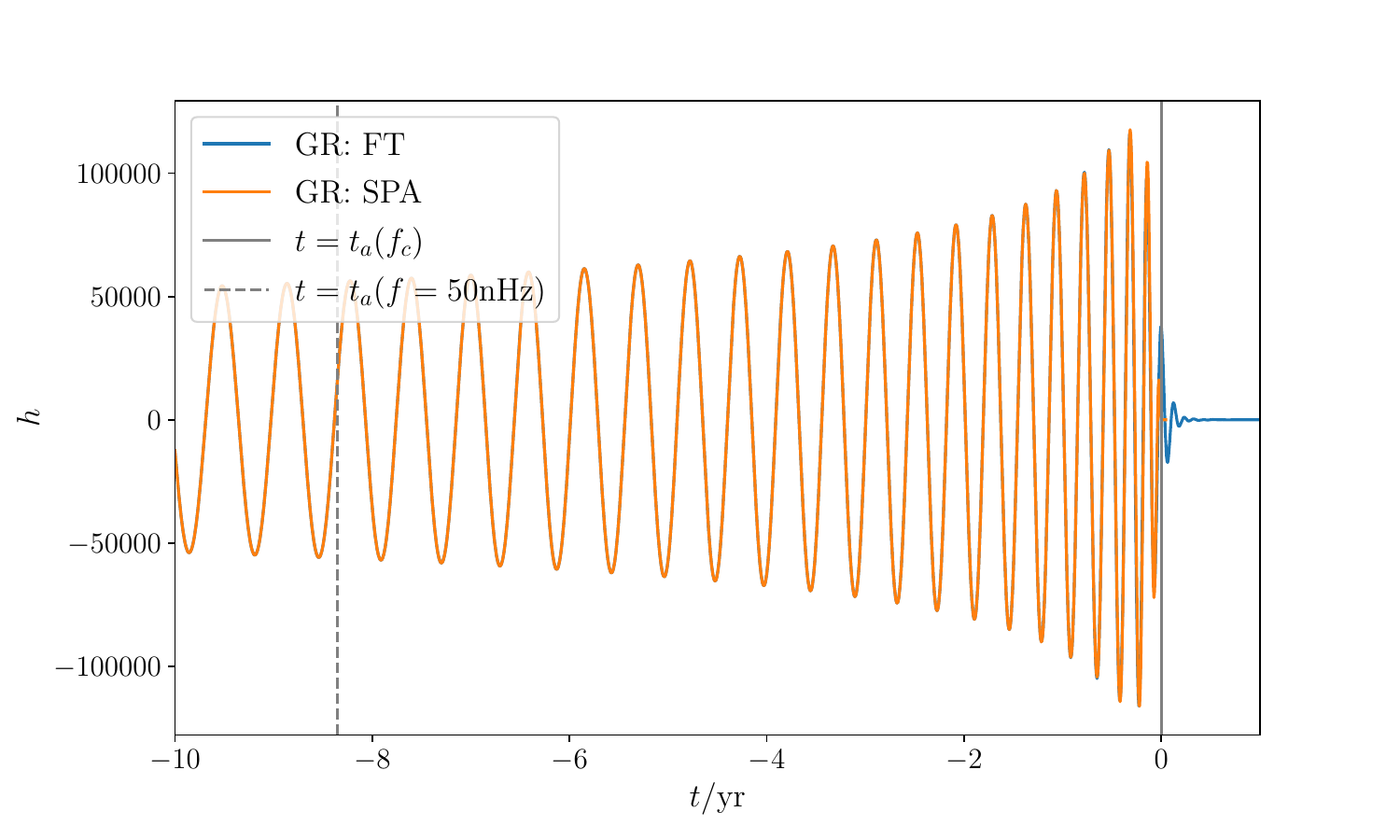}
    \caption{Time domain comparison between the stationary phase approximation and Fourier transform of the apodized binary black hole frequency domain GW waveform as shown in Fig.~\ref{fig:FD-hf}. 
    By convention, $t=0$ is the arrival time of the coalescence frequency $f_c$. The dashed vertical line indicates the arrival time of the typical PTA relevant frequency $f=50\,$nHz.
    }
    \label{fig:SPA-GR}
\end{figure}

We begin with small distances to the GW source, where propagation effects have not yet modified the waveform.   We can then apply the stationary phase approximation directly to the time domain expression Eq.~(\ref{eq:TD-ht}).
In this limit,
the frequency domain waveform takes the following form in the stationary phase approximation Eq.~\eqref{eq:SPA}, 
\begin{equation}\label{eq:FD-hf}
    \Tilde{h}(f) = \frac{Q}{D}(G\mathcal{M})^{5/6} f^{-7/6}\exp[i\Psi(f)] \ ,
\end{equation}
where $f\ge 0$, and
\begin{equation}
\label{eq,frequencydepphase}
    \Psi(f) = 2\pi f t_c - \phi_c -\frac{\pi}{4} +\frac{3}{4}(8\pi G\mathcal{M}f)^{-5/3} \ ,
\end{equation}
and as we shall see below more generally there is a modification to $\Psi(f)$ due to propagation over cosmological distances.

Note that following Appendix~\ref{app:SPA} with Eq.~\eqref{eq:FD-hf} and \eqref{eq,frequencydepphase}, one can verify that using the stationary phase approximation to perform the inverse Fourier transform recovers the time domain waveform Eq.~(\ref{eq:TD-ht}).
The derivative of the phase yields the arrival time of each frequency 
\begin{equation}
    t_a^{\rm GR}(f) = \frac{\partial\Psi}{\partial (2\pi f)} = t_c - 5 (8\pi f)^{-8/3} \left( G \mathcal{M} \right) ^{-5/3}  \ .
    \label{eq,taGR}
\end{equation}

Notice that the frequency in Eq.\ (\ref{eq:FD-hf}) extends unphysically to $f>f_c$, corresponding to $t \rightarrow t_c$ where $f_c=(6^{3/2}\pi G\mathcal{M})^{-1}$ is the coalescence frequency.
To obtain a realistic waveform, we 
use the Hann function to taper $\Tilde{h}(f)$ from its original value to zero between $(0.4f_c,f_c)$. Similarly, the Hann function taper also applies to the other end $(f_{\rm min},1.6f_{\rm min})$ where $f_{\rm min}=5\,$nHz is close to the lower limit of the detector frequency band.
We illustrate this apodized frequency domain waveform in Fig.~\ref{fig:FD-hf} with  a redshifted chirp mass $\mathcal{M}=1.3\times 10^{10}M_\odot$ corresponding to the maximal plausible SMBH mass\footnote{We obtain the chirp mass based on the assumption that the two black holes have same masses in this binary system.} \cite{Surti_2024} (chosen to maximize the frequency evolution), highlighting the position of $f_c$ and the typical PTA relevant frequency $50\,$nHz.

To explicitly verify the validity of the stationary phase approximation for this realistic waveform, we have numerically computed the time domain waveform from the apodized frequency domain waveform using both the SPA approach, following Eq.~(\ref{eq:SPA}), and the discrete Fourier transform. In Fig.~\ref{fig:SPA-GR}, we compare these different schemes demonstrating excellent agreement between them in the inspiral regime. The lack of agreement around the coalescence time is due to the remaining rapid frequency evolution after the arbitrary apodization which results in non-zero values for $t>0$ in the direct transform, while SPA predicts exactly zero in this region.
Notice that we choose the convention to make $t=0$ correspond to the arrival time of the coalescence frequency $f_c$, and this differs from $t_c$ which for the inspiral waveform is the arrival time of the unphysical $f\rightarrow \infty$ mode. We also show the arrival time of the typical PTA relevant frequency $f=50\,$nHz together with $f_c$.  
Notice that we have chosen $|t(f=50\,$nHz$)|$ 
to be comparable to the $T_{\rm obs} \sim$15-year PTA observation time window, again to maximize the frequency evolution. 
We will return to this example in \S~\ref{sec:example}, where a large frequency evolution corresponds to a large fractional deviation from the monochromatic overlap reduction function.

\begin{figure}   
\centering
\includegraphics[width=\textwidth]{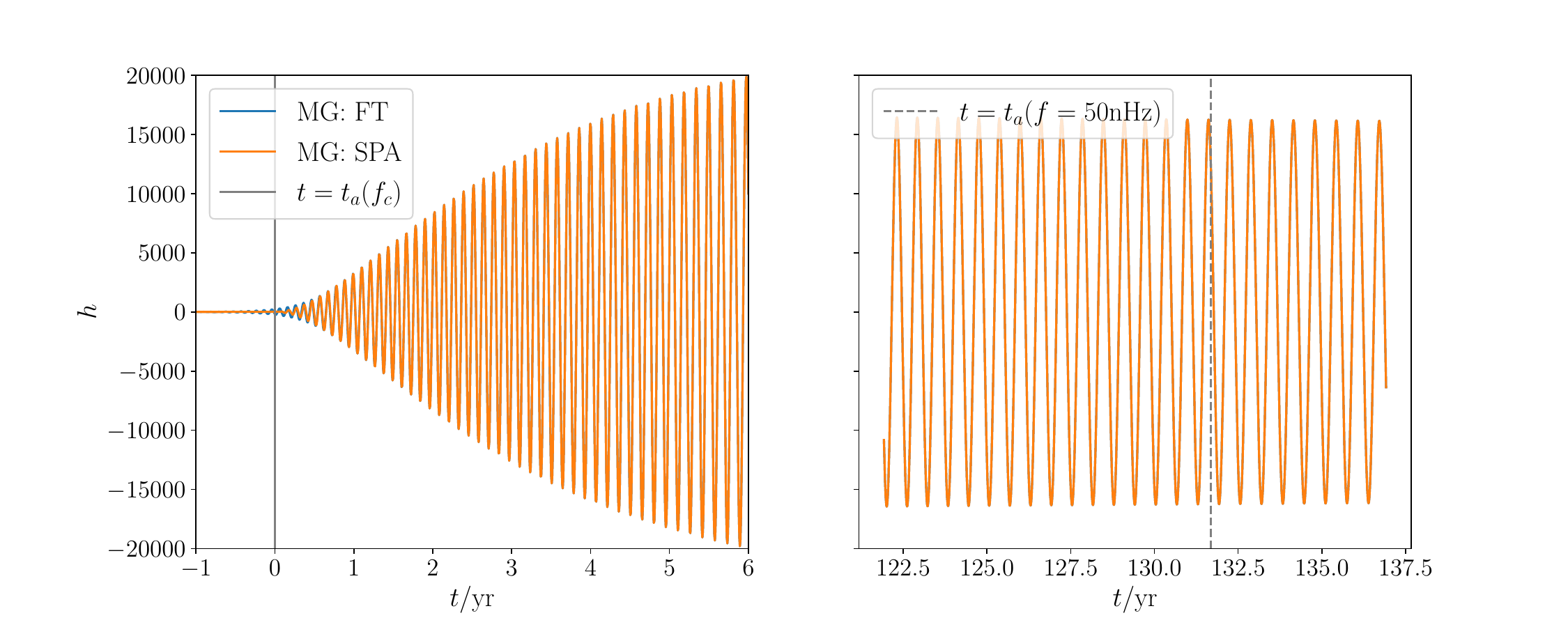}
    \caption{The time domain waveform at reception with modified gravity (here, massive gravity) corresponding to Figs.~\ref{fig:FD-hf} and \ref{fig:SPA-GR} for the  waveform in GR. 
    Here we have chosen $\mathcal{M}=1.3\times10^{10}M_{\rm sun}$, $D=8.77\,{\rm kpc}$, and $m_g=2.1\times 10^{-23}\,{\rm eV}$.
    Blue and orange curves indicate the predictions of the stationary phase approximation and superposition methods respectively.
    By convention, $t=0$ corresponds to the arrival time of coalescence frequency $f_c$ shown in the left panel. The typical PTA relevant frequency $f=50\,$nHz is shown in the right panel.  Note the gap in time that reflects the propagation time delay and reversal of the waveform.
    }
    \label{fig:SPA}
\end{figure}
 
In modified gravity (MG), there is a correction to the phase arising from the propagation  (see e.g.~\cite{Ezquiaga:2022nak})
\begin{equation}\label{eq:DPsi-mdr}
    \Delta \Psi(f) = - \int_0^{z_s} \frac{dz_s'}{H(z_s')}(2\pi f - k) \ ,
\end{equation}
that can be larger  than the emission term in 
Eq.~(\ref{eq,frequencydepphase}) for cosmological source redshifts.  Even if the frequency evolution is rapid in the emission waveform, the waveform at Earth may exhibit a much slower evolution and smaller deviations from a monochromatic overlap reduction function.

This phase shift is compatible with local propagation at the group velocity since $t_a(f)=\partial\Psi(f)/\partial (2\pi f)$ gives
\begin{equation}\label{eq:ta-mdr}
    t_a^{\rm MG}(f) =t_a^{\rm GR}(f)  + \int_0^{z_s} \frac{dz_s'}{H(z_s')} \left(\frac{1}{v_g(f,z_s')}-1\right) \ ,
\end{equation}
where the second term is the time delay due to propagation with MG.  

To make these considerations concrete, we focus on massive gravity as a specific example of modified gravity. The dispersion relation for massive gravity is
\begin{equation}
    \omega^2 = k^2 + \frac{m_g^2}{(1+z_s)^2}\ ,
\end{equation}
so that the group velocity $v_g \equiv \partial\omega/\partial k = k/\omega$. 
Using this, Eq.~\eqref{eq:DPsi-mdr}-\eqref{eq:ta-mdr} becomes
\begin{equation}\label{eq:DPsi-mg}
    \Delta \Psi(f) = - \int_0^{z_s} \frac{dz_s'}{H(z_s')}\left(2\pi f - \sqrt{(2\pi f)^2 - \frac{m_g^2}{(1+z_s)^2}} \right) \ ,
\end{equation}
yielding
\begin{equation}\label{eq:ta-mg}
    t_a^{\rm MG}(f) = 
        t_a^{\rm GR}(f) + \int_0^{z_s} \frac{dz_s'}{H(z_s')} \left(\frac{ 2\pi f  }{\sqrt{4\pi^2 f^2 - m_g^2/(1+z_s')^2 }} -1\right) \ .
\end{equation}

Following the same procedure as before, but now with this modified phase and arrival time, we can apply the SPA  for the transformation from the frequency to the time domain in MG.  
Since modes where $f< m_g/2\pi$ do not propagate, for the frequency domain apodization, we have $f_{\rm min} \rightarrow {\rm max}(m_g/2\pi, f_{\rm min})$ and will choose $m_g = 2\pi\times5\,$nHz $\approx 2.1 \times 10^{-23}$eV as an example. 

We compare the SPA prediction to the superposition results
for a relatively small distance $D=8.77\,$kpc, comparable to our nearest supermassive black hole at the center of the Galaxy, in Fig.~\ref{fig:SPA}, demonstrating that SPA remains valid at reception in this case and can be used to the arrival times of various frequency components of the waveform.  
Notice the much larger and inverted temporal separation between the arrival of $f_c$ (left panel) and $f=50\,$nHz (right panel) as compared with Fig.~\ref{fig:SPA-GR}.
Since lower frequencies propagate slower, the latter arrives after the former, despite being emitted earlier.
The amplitude of the time domain GW also decreases despite the frequency domain amplitude remaining the same because the same emission power is now spread out over a longer arrival time.  
In the SPA this can be directly computed from Eq.~(\ref{eq:ta-mg}) for the modified arrival times
\begin{equation}
\label{eq:amplitudescaling}
    \left( \frac{h_{\rm MG}(t(f))}{h_{\rm GR}(t(f))} \right)^2 = \frac{\partial t_{\rm GR}/\partial f}{\partial t_{\rm MG}/\partial f} \ .
\end{equation}
We can quantify these effects more directly in the limit $f \gg m_g/2\pi$ which is a good approximation in the displayed region of Fig.~\ref{fig:SPA-GR}.   Here
\begin{equation}\label{eq:ta-MGlinear}
 \Delta t_a \equiv   t_a^{\rm MG}-t_a^{\rm GR} \approx \frac{m_g^2}{8\pi^2f^2} D_{\rm eff} \ ,
\end{equation}
with
\begin{equation}
\label{eq,Deff}
    D_{\rm eff} = \int_0^{z_s}\frac{dz'_s}{H(z'_s) (1+z'_s)^2} \ ,
\end{equation}
where the extra factor of $(1+z_s')^2$ with respect to $D$ comes from the redshifting of the GW frequency with respect to $m_g$ which reduces the propagation delay at high redshift.
For a given frequency, there is an effective distance where $t_a^{\rm MG}(f)-t_c=0$, 
\begin{equation}
\label{eq,Dcross}
D_{\rm cross} = \frac{5}{32 \pi^{2/3}} f^{-2/3} (G {\cal M})^{-5/3} m_g^{-2}.
\end{equation}
For $D_{\rm eff}>D_{\rm cross}$ this frequency is received after $t_c$ despite being emitted before coalescence, effectively time reversing the waveform.  Note that $D_{\rm cross}$ increases at lower frequencies implying that at a fixed $D_{\rm eff}$, sufficiently low frequencies arrive in the normal time ordering, but recall that we have a lower limit on the propagation frequency such that $f\gg m/2\pi$ in the expansion of $\Delta t_a$. 

We can also derive the amplitude change implied by Eq.~(\ref{eq:amplitudescaling}) and (\ref{eq:ta-MGlinear})   for the large $D_{\rm eff}$ case where  $\partial t_{\rm MG}/\partial f \propto D f^{-3}$. Since
$\partial t_{\rm GR}/\partial f \propto f^{-11/3}$ and $h_{\rm GR}(t(f)) \propto (t(f))^{-1/4} \propto f^{2/3}$,
the  amplitude scales as $h_{\rm MG}(t(f)) \propto  f^{1/3} D^{-1/2}$, which can be very small for large distances but is actually a {\it less} steep frequency dependence or ``chirp" than in GR.  Correspondingly in Fig.~\ref{fig:SPA}, the effect of apodization is much more pronounced around the $t=0$ arrival time of $f_c$ compared to the chirp in Fig.~\ref{fig:SPA-GR}.

More generally, this propagation time delay between frequency components is the result of group velocity dispersion, i.e., the frequency dependence of $v_g(f)$. However, unlike the Gaussian case, the manifestation of this waveform distortion is to spread out the arrival times of neighboring frequencies and hence to make the wave {\it more monochromatic} over a given observation time. In particular around the arrival time of $f=50\,$nHz in Fig.~\ref{fig:SPA}, the arrival frequency evolves negligibly across the 15yr PTA baseline. We shall now see that when this propagation time delay dominates, the overlap reduction function become closer to the simple monochromatic form given in \S~\ref{sec:mono}.

\subsection{Cross-Frequency Overlap Reduction Function}
\label{sec:example}

Having verified the SPA for the inspiral waveform, we can now use the stationary phase frequency as a function of arrival time, $f(t)$, to assess the modifications to the overlap reduction function. Even though at each arrival time the redshift induced by the GW follows the monochromatic form of Eq.~(\ref{eq,monoz}) to a good approximation, the correlated residuals at two different arrival times involve a cross-correlation between two different frequencies. To the extent to which the arrival frequency evolves over the temporal baseline of the observation time, $T_{\rm obs}$, the overlap reduction function is also modified from its monochromatic form. Specifically, the modification of
Eq.~(\ref{eq:RR-mono}) becomes 
\begin{equation}\label{eq:R1R2}
    R(t_1,\hat{p}_1)R^*(t_2,\hat{p}_2)  = \frac{H\left(f_1(t_1),\hat{p}_1\right) H^*\left(f_2(t_2),\hat{p}_2\right)} {(2\pi)^2f_1(t_1)f_2(t_2)} \frac{\hat p_1^i \hat p_1^j e_{ij}^A}{2}\frac{\hat p_2^k \hat p_2^l e_{kl}^B}{2}  h_0^A(t_1) h_0^{B*} (t_2) \ .
\end{equation}
Hence, the overlap reduction function  reads
\begin{equation}
\label{eq,angularcorr_cross}
  \Gamma(\xi,t_1,t_2)_{\rm cross} 
  \equiv \frac{1}{2} \frac{ \sum_{\ell=2}^\infty \frac{2\ell+1}{4\pi} C_\ell(t_1,t_2) P_\ell(\cos\xi)} {\sum_{\ell=2}^\infty \frac{2\ell+1}{4\pi} C_\ell} \ ,
\end{equation}  
with the total power spectrum 
\begin{eqnarray}
    C_\ell(t_1,t_2) &=& \frac{1}{2 \ell+1} \sum_m a_{\ell m}(t_1) a^*_{\ell m}(t_2) +\rm h.c. \nonumber\\
    &=& \frac{\pi}{8} \frac{c_\ell(v_p(t_1))c_\ell^* (v_p(t_2)) }{(2\pi)^2 f_1(t_1)f_2(t_2)} \frac{(\ell-2)!}{(\ell+2)!}\langle h_0^+(t_1)h_0^{+*}(t_2) + h_0^\times(t_1)h_0^{\times*} (t_2) \rangle +\rm h.c. \ ,
\end{eqnarray}
where $c_\ell(v_p)$ is the coefficient defined in Eq.\ \eqref{eq,cell}, and $v_p(t_{1,2})$ are the phase velocities of the corresponding arrival frequencies, $f_{1,2}$, arrive at times $t_{1,2}$, respectively.
Notice that the gravitational wave amplitude $h_0^{+,\times}$ again drops out of the ORF. When $t_1=t_2$,  or more specifically, $f_1 = f_2$, the result reduces to the monochromatic case Eq.\eqref{eq,angularcorr}.
Deviations of $\Gamma_{\rm cross} $ from the monochromatic result, $\Gamma_m$, therefore depend on the arrival frequency evolution across the timing baseline.

\subsubsection{Nearby source}
For pegagogical purposes,
we begin by considering a source located very close to Earth $D\rightarrow 0$ with an extremely massive GW source with $\mathcal{M}=1.3\times 10^{10}M_\odot$ corresponding to the GW waveform of Fig.~\ref{fig:SPA-GR}.  Notice that the frequency evolves significantly over a $T_{\rm obs} \sim 15\,$yr observational time due to the proximity of $50$nHz to the coalescence frequency with such a high chirp mass. 

\begin{figure}   
\centering
\includegraphics[width=0.49\textwidth]{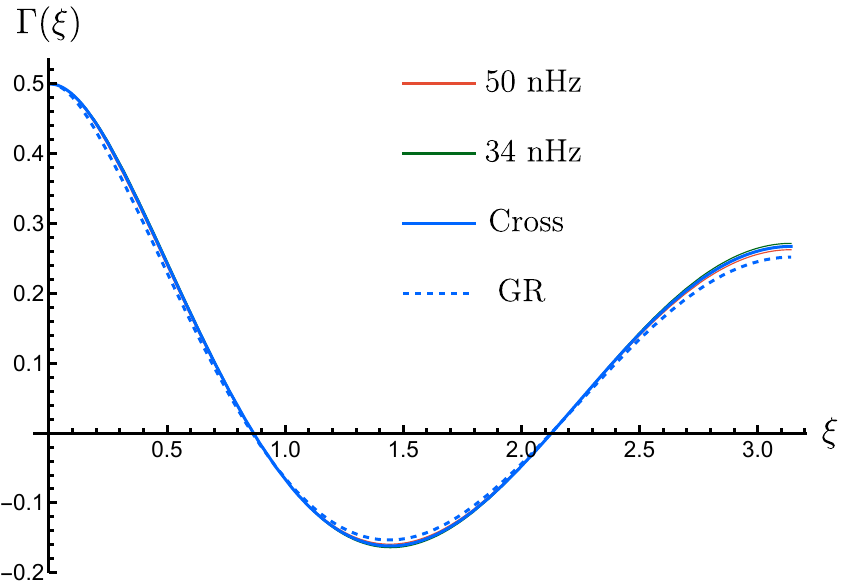} 
\includegraphics[width=0.49\textwidth]{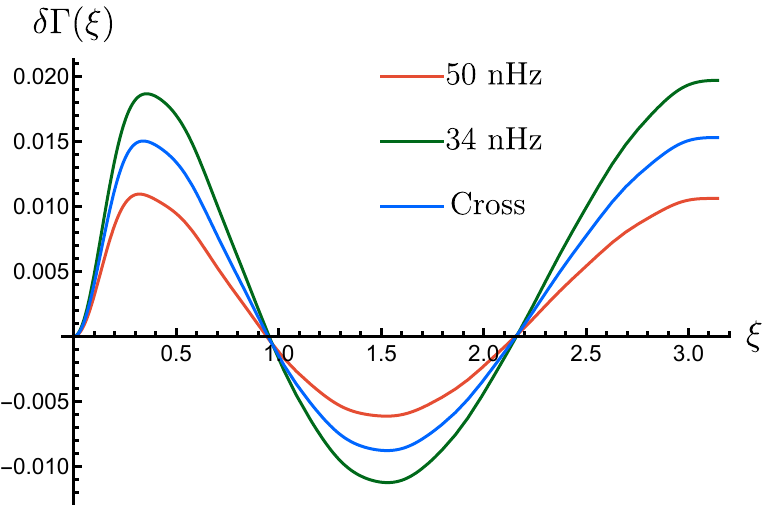}
    \caption{
    {\it Left panel:} dashed blue line: the Hellings-Downs ORF in GR; red and green lines: the monochromatic ORF defined in Eq.~\eqref{eq,angularcorr} at  $f= 50\,$nHz ($v_p = 1.005$) and $f= 34\,$nHz ($v_p = 1.011$) respectively;  solid blue line: the  ORF with cross-frequency correlation, ($f_1 = 34\,$nHz and $f_2 = 50\,$nHz). Notice the cross-frequency correlation is very close to the monochromatic ORF with the mean frequency (see text). 
    {\it Right panel:} deviation from the Hellings-Downs curve with the same parameters as the left panel. The amplitude of the difference compared to HD curve can be well estimated by the end point $\xi=\pi$. 
    }
    \label{fig:unequaltimecorr}
\end{figure}

Specifically we take two frequencies that are separated in arrival time by
\begin{eqnarray}
     t_a(f_2=50\,{\rm nHz}) - t_a(f_1=34{\rm nHz} ) \approx 15\,{\rm years}\ .
\end{eqnarray}
We also maximize the deviations from GR by taking a large graviton mass
$m_g/2\pi = 5\,{\rm nHz}$.
For this parameter choice,  the phase velocities at the two frequencies are 
\begin{eqnarray}
    v_p(34\,\text{nHz}) = 1.011\ , \quad v_p(50\,\text{nHz} ) = 1.005\ .
\end{eqnarray}
In Fig.\ \ref{fig:unequaltimecorr}, we show these corresponding monochromatic predictions for the whole overlap reduction function $\Gamma(\xi)$ and their deviations from GR,  $\delta\Gamma = \Gamma-\Gamma_{\rm GR}$.   
We also show the cross frequency prediction in Fig.\ \ref{fig:unequaltimecorr}.   Notice that the change from the two monochromatic results is a substantial fraction of the net deviation from GR, and therefore the frequency evolution is important to include. 
 More quantitatively for the two monochromatic results, $\delta \Gamma_m(\pi)|_{f_1}=0.012$ and
$\delta \Gamma_m(\pi)|_{f_2}=0.021$, which substantially differ. 
On the other hand, the cross result should be very close to the mean of these two.  
Since the phase velocity is close to the speed of light, we can understand this behavior analytically. Notice that
$v_p-1 =\varepsilon \propto f^{-2}$ and the deviation $\delta \Gamma_m$ in Eq.\ \eqref{eq,deltaGamma} is linear in this quantity up to logarithmic corrections.  The linear correction in the product of $c_\ell(v_p)$ implies that the combination of the two frequencies is well approximated by the monochromatic result at the inverse variance mean value,
\begin{eqnarray}
\label{eq,meanfrequency}
    \bar f = \sqrt{\frac{2 f_1^2 f_2^2}{f_1^2+  f_2^2}}\ .
\end{eqnarray}
In our example the predicted
$\delta \Gamma_m(\pi)|_{\bar f} = 0.0172$ whereas the cross frequency result is 
$\delta \Gamma_{\rm cross}(\pi)\equiv \Gamma(\pi)_{f_1,f_2}
-\Gamma_{\rm GR}(\pi) = 0.0171$. Using the mean frequency accounts for the deviations, $\delta \Gamma_m(\pi)|_{f_2}-\delta \Gamma_m(\pi)|_{f_1} = 0.009$, that are linear in $\varepsilon  \Delta \ln f$, where $\Delta \ln f = \ln f_2 -\ln f_1$,
and leaves the next leading order $\delta \Gamma_{\rm cross} - \delta \Gamma_m|_{\bar f} \sim \mathcal{O}(\varepsilon^2\Delta \ln f)$,
due to the expansion of $c_\ell$ when $|\Delta\ln f|\ll 1$.  In this case where $\Delta\ln f =0.38$ is order unity, the correction is ${\cal O}(\varepsilon^2)$ and since $\varepsilon_{\bar f}| = 0.008$, the residual $\delta\Gamma_{\rm cross} - \delta \Gamma_m|_{\bar f} \approx 10^{-4}$ scales as expected.

Therefore for other parameter choices the main considerations are the values of $\varepsilon$, the deviation in the phase velocity, and $\Delta \ln f$ - the frequency evolution between the observations.  While the difference in the full cross ORF from GR scales as $\varepsilon \Delta\ln f$, the difference between the cross and monochromatic  at the mean frequency scales as $\varepsilon^2 \Delta\ln f$.

\subsubsection{Distant source}
We now turn to the opposite extreme, where the source is located at a large distance from Earth, so that the second term of Eq.~\eqref{eq:ta-mg}, namely the delay due to the modified dispersion relation, dominates. In the $m_g/2\pi\ll f$ limit, 
this corresponds to an effective distance $D_{\rm eff} \gg D_{\rm cross} \propto f^{-2/3}{\cal M}^{-5/3} m_g^{-2}$ (see Eq.~\eqref{eq,Deff} and \eqref{eq,Dcross}).

In this limit, the arrival time can be approximated by the propagation delay Eq.~\eqref{eq:ta-MGlinear}. The change in the arrival frequency across the observation baseline is
\begin{equation}\label{eq:Dlnf-MGlinear}
    \Delta\ln f \approx \bigg|\frac{\partial \ln f}{\partial t_a}\bigg|T_{\rm obs} \approx \bigg|\frac{\partial \ln f}{\partial \Delta t_a}\bigg|T_{\rm obs} \approx  \frac{T_{\rm obs}}{2\Delta t_a}  \propto m_g^{-2} D_{\rm eff}^{-1} \ .
\end{equation}

In the example shown in Fig.~\ref{fig:SPA} with $m_g=2\pi\times 5\,{\rm nHz}\approx 2.1\times10^{-23}\,{\rm eV}$, and $D_{\rm eff}\approx D =8.77{\rm kpc}$, we have $f_2 = 50\, $nHz and $\Delta \ln f\approx0.056$ for the same $T_{\rm obs} \approx 15 \rm yr$. The mean frequency can be obtained from Eq.\eqref{eq,meanfrequency}, $\bar f=48.5\,$nHz. Notice that the chirp mass drops out of frequency evolution\ (\ref{eq:Dlnf-MGlinear}) in this $D_{\rm eff}\gg D_{\rm cross}$ limit, so our example applies to any ${\cal M}$.

The total deviation from GR in the cross frequency ORF, 
$\delta \Gamma_{\rm cross}(\pi) = 0.0128$,
remains comparable to the $D \rightarrow 0$ case since $\varepsilon_{  {\bar f} } \approx 0.005$ is independent on distance. On the other hand, the frequency evolution is an order of magnitude smaller, $\delta\Gamma_{m}|_{  f_1}-\delta\Gamma_{m}|_{ f_2} \approx 0.001$, which scales as $\varepsilon \Delta\ln f$ as expected.
Moreover this small evolution is well captured by evaluation of the monochromatic deviation at the mean $\bar f$:
$\delta \Gamma_{\rm cross}(\pi)-\delta \Gamma_m(\pi)|_{\bar f} \approx 10^{-6}$. This is consistent with our scaling expectation $\varepsilon^2\Delta\ln f$.

Scaling this distance to cosmological scales $\sim 1\,$Gpc using Eq.~(\ref{eq:Dlnf-MGlinear}) makes this correction extremely small at this large graviton mass. At smaller $m_g$, $\Delta \ln f$ and the change due to frequency evolution can be larger but the overall deviation in the ORF from GR is also smaller (See Eq.~\eqref{eq,deltaGamma} with $\varepsilon=v_p-1\approx m_g^2/(8\pi^2f^2)$ in this limit). 
Lowering the chirp mass ${\cal M}$ raises the distance $D_{\rm cross}$ at which the propagation time delay dominates, but correspondingly decreases the frequency evolution as $\Delta \ln f \propto {\cal M}^{5/3}$ if $D_{\rm eff} \ll D_{\rm cross}$.
In principle the frequency can evolve faster or be double valued for $D_{\rm eff} \sim D_{\rm cross}$, or more specifically where $\partial t_a/\partial\ln f=0$. However, this scenario would require fine-tuning that is only feasible for single-source events and cannot be consistently maintained across a population of sources.

\section{Conclusion and Discussion}\label{sec:conclude}

In this paper we have considered the implications of relaxing the assumption of a stochastic gravitational wave background consisting of uncorrelated monochromatic plane waves in the analysis of timing residuals from PTAs. 
While there are good reasons for assuming a stochastic background behaves in this manner, if the background is composed of unresolved astrophysical sources then in principle there could be a markedly different correlated signal. 
Moreover, gravitational waves from sources propagate causally as wavepackets which result in differences if there is a modification to General Relativity involving their dispersion relation that is not just characterized by the phase of monochromatic components.  

To understand the GW propagation effects on PTA observations, we began by exploring the simple thought example of a Gaussian wavepacket. Here we highlighted and resolved the apparent paradox that the wavepacket travels at the group velocity whereas the monochromatic results depend only on the phase velocity at the Earth and the pulsars.  Using both a propagation approach and a superposition approach at the Earth and pulsars, we showed that the group velocity enters into the amplitude and stationary phase value of the Earth and pulsar terms and effectively averages the monochromatic result over the frequency distribution of the wavepacket.  Furthermore, the wavepacket distorts during propagation if the group velocity dispersion becomes significant across the long distance to the gravitational wave source. 
As expected, when the Gaussian wavepacket is narrow in the frequency domain, the correction to the PTA response remains minimal, even in the context of modified gravity theories.

Building on this understanding of wavepacket effects, we considered realistic waveforms corresponding to  Supermassive Black Hole Binary inspiral signals of nearby and distant sources.  We found that the frequency evolution of the chirp together with the frequency-dependent delays of the modified dispersion relation induces cross-frequency correlations in the PTA signals, changing the overlap reduction function (ORF). 
However, for sources at cosmological distances and the dispersion relation of massive gravity, the long time delays between different frequencies implied by the group velocity dispersion distorts the wavepacket making it  closer to monochromatic across a realistic PTA temporal baseline of $\sim 15$yrs.  When the frequency evolution across this baseline is small, the induced cross-frequency ORF can be well estimated by the monochromatic result at the mean frequency of two arrival frequencies. 

Therefore, for these astrophysical sources at cosmological distances, the monochromatic assumption remains valid for analysis and can be effectively used to constrain deviations from GR. Nevertheless, in nearby single-source events, it is possible in the future \cite{NANOGrav:2023pdq} to detect the influence of these waveform effects, offering new opportunities to test GR.

For other cosmological sources such as primordial gravitational waves, or those originating from phase transition in the early universe, the uncorrelated monochromatic assumption should be even more valid.  For other binary sources, such as primordial black hole binaries, if they merge at high redshift, the resulting frequency will be redshifted to the nanohertz band and may also contribute to the SGWB. For these more general sources we have provided scaling relations for the overlap reduction function given the frequency evolution and the amplitude of the phase velocity differences, to assess the extent to which these effects could be important in tests of GR.

\acknowledgements
We thank Austin Joyce and Misao Sasaki for interesting discussions.
W.H. is supported by U.S. Dept. of Energy contract DE-FG02-13ER41958 and the Simons Foundation. Q.L. is supported by World Premier International Research Center Initiative (WPI), MEXT, Japan. M.X.L. is supported by funds provided by the Center for Particle Cosmology at the University of Pennsylvania. 
The work of M.T. is supported in part by DOE (HEP) Award DE-SC0013528.

\appendix

\section{Stationary phase approximation}\label{app:SPA} 

This section serves as a reminder of the standard stationary phase approximation technique. The stationary phase approximation is used to evaluate oscillatory integrals under the assumption that the phase changes slowly with respect to the variable of integration. This technique can be applied to both Fourier and inverse Fourier transforms. For illustration, we begin with the Fourier domain; the inverse process can be achieved by simply changing the variables.

For a given Fourier domain waveform with a known phase function
\begin{equation}
    \tilde{h}(f) = |\tilde{h}(f)|e^{i\Psi(f)} \ ,
\end{equation}
the time domain waveform can be obtained through an inverse Fourier transform,
\begin{equation}
    h(t) \equiv \int_{-\infty}^{\infty} e^{- 2 \pi i f t} \tilde{h}(f) d f\ .
\end{equation}
We find the stationary phase point through,
\begin{equation}
\label{eq:SPP}
    \frac{d\Psi }{df}\bigg|_{f_0} = 2\pi t\ ,
\end{equation}
and then expand the phase around this value as
\begin{equation}
    \Psi(f)\approx \Psi(f_0)+ (f-f_0) \Psi'(f_0)+ \frac{1}{2}(f-f_0)^2 \Psi''(f_0) \ .
\end{equation}
The waveform can therefore be approximated as
\begin{eqnarray}\label{eq:SPA}
  h(t) &\approx&  |\tilde{h}(f_0)| e^{  i \Psi(f_0)-2\pi i f_0 t  } \int_{-\infty}^{\infty} e^{\frac{i}{2}(f-f_0)^2 \Psi''(f_0)}    d f \nonumber\\
   &=& |\tilde{h}(f_0)|  \sqrt{\frac{ 2 \pi}{|\Psi''(f_0)|}}  e^{ i [\Psi(f_0)-2\pi  f_0 t  \pm  \frac{\pi}{4}] }   \ ,    
\end{eqnarray}
where $\pm$ corresponds to positive (negative) $\Psi''(f_0)$ respectively, and we have used the Gaussian integral 
\begin{equation}
    \int_0^{\infty} e^{i a x^2} d x=e^{i \pi \operatorname{sgn}(a) / 4} \sqrt{\frac{\pi}{4|a|}}
\end{equation}
in the final step.

\section{Coefficients of redshift expansion}\label{app:Bn}

Here we provide details of the derivation of the coefficients in the redshift expansion in Eq.~\eqref{eq,zpropagation}.
Combining Eq.~\eqref{eq,superposition} and \eqref{eq:hA-superposition}, we may express $I_e$ as,
\begin{eqnarray}
  I_e & = &\sqrt{2\pi} \sigma_x h_0^A \int \frac{dk}{2\pi} H(k)e^{-\frac{(k-\bar k)^2\sigma_x^2}{2}} e^{i\varphi_e(k)}   \nonumber\\
  &\approx & \sqrt{2\pi} \sigma_x h_0^A e^{i\bar \phi_e} \int \frac{dk}{2\pi} H(k)e^{-\frac{(k-\bar k)^2\sigma_x^2}{2}} e^{ i \varphi_e'(\bar k)(k-\bar k)} \ ,
\end{eqnarray}
where we have neglected the dispersion of the group velocity, $d^n \varphi_e/dk^n|_{k=\bar k} =0$. 

We can further Taylor expand $H(k)$ around $k=\bar k$, and use the saddle point approximation to express the integral as a sum of a series of Gaussian integrals,
\begin{eqnarray}
    I_e &\approx& \sqrt{2\pi} \sigma_x h_0^A e^{i\bar \phi_e} \sum_{n=0} \frac{d^n H(k)}{n! dk^n}\bigg|_{k=\bar k} \int \frac{dk}{2\pi} (k-\bar k)^n e^{-\frac{(k-\bar k)^2\sigma_x^2}{2}+ i \chi_e \sigma_x (k-\bar k)} \nonumber\\
    &=& \sqrt{2\pi} \sigma_x h_0^A e^{i\bar \phi_e -\frac{\chi_e^2}{2}} \sum_{n=0}^\infty \frac{ H^{(n)}(\bar k)}{n!} \int \frac{d\Delta k}{2\pi} \left(\Delta k + i\frac{\chi_e}{\sigma_x}\right)^n e^{-\frac{\Delta k^2\sigma_x^2}{2} } \ , 
\end{eqnarray}
where $\Delta k = k-\bar k - i\frac{\chi_e}{\sigma_x}$, and $H^{(n)}(\bar k)$ represents the n-th derivative of $H(k)$ evaluated at $k = \bar k$. 

Using the binomial expansion, 
\begin{eqnarray}
    \left(\Delta k + i\frac{\chi_e}{\sigma_x}\right)^n =  \sum_{m=0}^n  \frac{n!}{m!(n-m)!} \left( i\frac{\chi_e}{\sigma_x}\right)^{n-m} \Delta k^m  \ ,
\end{eqnarray}
and the Gaussian integral
\begin{eqnarray}
   \int \frac{d\Delta k}{2\pi}\Delta k^m e^{-\frac{\Delta k^2\sigma_x^2}{2} } = \begin{cases}
      \frac{1}{2\pi} \Gamma\left(\frac{m+1}{2}\right) \left(\frac{\sigma_x^2}{2}\right)^{-\frac{m+1}{2}}  & \text{for even } m\\
      0 & \text{for odd } m
    \end{cases}  \ ,
\end{eqnarray}
we can express the coefficients $B_{n,e}$ as,
\begin{eqnarray}
    B_{n,e} &=&  \frac{\sigma_x}{\sqrt{2\pi}} \sigma_x^n   \sum_{m=0, \text{even}}^n  \frac{n!}{m!(n-m)!} \left( i\frac{\chi_e}{\sigma_x}\right)^{n-m}  \Gamma\left(\frac{m+1}{2}\right) \left(\frac{\sigma_x^2}{2}\right)^{-\frac{m+1}{2}} \nonumber\\
    &=&   i^n \chi_e 2^{\frac{n-1}{2}}    U\left(\frac{1-n}{2}, \frac{3}{2}, \frac{\chi_e^2}{2}\right)\ ,
\end{eqnarray}
where $U(a,b,z)$ is the confluent hypergeometric function. 
This gives the results in Eq.~\eqref{eq:Bn}.

\bibliography{ref} 
\end{document}